\documentstyle[graphicx,aps,multicol]{revtex}

\newcommand{\eq}{\begin{equation}}
\newcommand{\en}{\end{equation}}
\newcommand{\eqa}{\begin{eqnarray}}
\newcommand{\ena}{\end{eqnarray}}
\newcommand{\ml}{\begin{mathletters}}
\newcommand{\eml}{\end{mathletters}}

\begin{document}


\title{Temporal correlations and neural spike train entropy}

\author{Simon R. Schultz$^1$ and Stefano Panzeri$^2$}
\address{$^1$ Howard Hughes Medical Institute and Center for Neural
Science, New York University,\\ 4 Washington Place, New York, NY
10003, U.S.A.\\
$^2$ Department of Psychology, Ridley Building, University of
Newcastle upon Tyne, \\
Newcastle upon Tyne, NE1 7RU, U.K.
}


\maketitle

\begin{abstract}
Sampling considerations limit the experimental conditions under
which information theoretic analyses of neurophysiological data
yield reliable results. We develop a procedure for computing the
full temporal entropy and information of ensembles of neural spike
trains, which performs reliably for limited samples of data. This
approach also yields insight upon the role of correlations between
spikes in temporal coding mechanisms. The method, when applied to
recordings from complex cells of the monkey primary visual cortex,
results in lower RMS error information estimates in comparison to
a `brute force' approach. \vspace{0.2cm} \noindent PACS numbers:
87.19.Nn,87.19.La,89.70.+c,07.05.Kf
\end{abstract}

\begin{multicols}{2}

\begin{figure}
\narrowtext
\includegraphics[height=4cm, width=8cm]{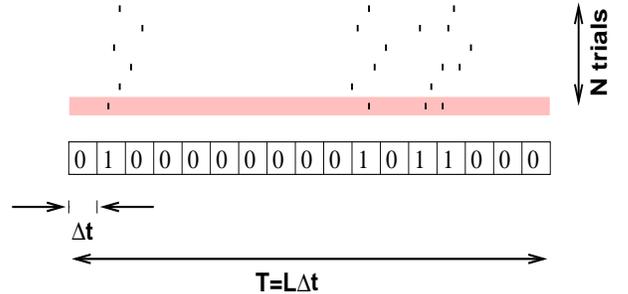}
\caption{Digitising spike trains into binary `words' with a given
precision. A common experimental structure has $N$ repeats for each
separate stimulus (one stimulus shown). The spike emission times for
each such `trial' are binned with resolution $\Delta t$, as shown for
the last raster. There are $2^L$ possible words when examining data
from a time window of duration $T$.}
\label{fig:binning}
\end{figure}

Cells in the central nervous system communicate by means of
stereotypical electrical pulses called action potentials, or spikes
\cite{Adr26}. The Shannon information content of neural spike trains
is fully described by the sequence of times of spike emission. In
principle, the pattern of spike times provides a large capacity for
conveying information beyond that due to the code commonly assumed by
physiologists, the number of spikes fired
\cite{Mac+52}. Reliable quantification of this spike timing
information is made difficult by undersampling problems that scale
with the number of possible spike patterns, and thus up to
exponentially with the precision of spike observation (see
Fig.~\ref{fig:binning}). While advances have been made in experimental
preparations where extensive sampling may be undertaken
\cite{The+96,deR+97,Rie+97,Strong98}, our understanding of the
temporal information properties of nerve cells from less
accessible preparations such as the mammalian cerebral cortex is
limited.

Any direct estimate of the complete spike train information is limited
by sampling considerations to relatively small wordlengths, and
therefore to the analysis of short time windows of data. However, it
is possible to take advantage of this restriction itself to obtain
estimators which have better sampling properties than a `brute
force' approach. In this Letter we present an approach based upon a
Taylor series expansion of the entropy, to second order in the time
window of observation \cite{serieslit}. The analytical expression so
derived allows the ensemble spike train entropy to be computed from
limited data samples, and relates the entropy and information to the
instantaneous probability of spike occurrence and the temporal
correlations between spikes. Comparison with other procedures such as
the `brute force' approach \cite{deR+97,Bur+98} indicates that our
analytical expression gives substantially better performance for data
sizes of the order typically obtained from mammalian neurophysiology
experiments, as well as providing insight into potential coding
mechanisms.

Consider a time period of duration $T$, associated with a dynamic or
static sensory stimulus, during which the activity of $C$ cells is
observed. The neuronal population response to the stimulus is
described by the collection of spike arrival times $\{t^a_i\}$,
$t^a_i$ being the time of the $i$-th spike emitted by the $a$-th
neuron. The spike time is observed with finite precision $\Delta t$,
and this bin width is used to digitise the spike train
(Fig.~\ref{fig:binning}). For a given discretisation (temporal
precision), the entropy of the spike train is a well defined
quantity. The total entropy of the spike train ensemble is
\eq H(\{t^a_i\}) = - \sum_{\{t_i^a\}} \;
P(\{t_i^a\})\log_2 P(\{t_i^a\}) , \label{eq:entropy}\en where the summation
is over all possible spike times within $T$ and over all
possible total spike counts from the population of cells. This entropy
quantifies the total variability of the spike train. Each different
stimulus history (time course of characteristics within $T$) is
denoted as $s$. The noise entropy, which quantifies the variability to
repeated presentations of the same stimulus, is $H^{\rm noise} =
\left< H(\{t_i^a\}|s) \right>_s$, where the angular brackets indicate
the average over different stimuli, $\left< A(s) \right>_s \equiv
\sum_{s\in \cal S} P(s) A(s)$. The mutual information that the
responses convey about which stimulus history invoked the spike train
is the difference between these two quantities.

\vspace{-0.3cm}

These entropies may be expanded as a Taylor series in the time window
of measurement,
\eq
H = T H_t + \frac{T^2}{2} H_{tt} + O(T^3) .
\en
\end{multicols}
\widetext

To compute the Taylor expansion, we made the following assumptions:
(i) The time window is short enough, or the firing rate low enough,
that there are few spikes per stimulus presentation. (ii) The entropy
is analytic in $T$. (iii) Different trials are random realisations of
the same process. We will use the bar notation for the average over
trials at fixed stimulus, such that if $r_a(t; s) = \sum_i
\delta_{t,t_i^a(s)}$, the time-dependent instantaneous firing
rate $\overline{r}_{a}(t;s)$ is its average over experimental
trials. (iv) Spikes are not locked to each other with infinite
precision; in other words, the conditional probability of a spike
occuring at time $\tau_j^b$ given occurrence of a particular spike
pattern $\{t^a_i\}$ scales for small $\Delta t$ proportionally to
$\Delta t$ plus higher order terms, with {\em no} $O(1)$ terms:
$P(\tau_j^b|\{t^a_i\};s) \propto \Delta t + \cdots$ for each possible
spike pattern $\{t^a_i\}$. The validity of these assumptions has been
examined elsewhere
\cite{tempinfo}.

The probability of observing a pattern with $k$ spikes can be
expressed as a product of $k$ probabilities of each of the spikes
given the presence of others. Thus from (iv), the probability of this
pattern is proportional to $\Delta t^k$, and the expansion is
essentially in the total number of spikes emitted. This also implies
that only the conditional probabilities between spike pairs are
necessary for the 2nd order expansion. Parameterising the conditional
probability between two spikes by the scaled correlation
$\gamma_{ab}(t^a_i,t^b_j;s)$
\cite{ref:gamma}, we can now write down the probabilities
required by Eq.~1.

Denoting the {\em no spikes} event as {\bf 0} and the joint occurrence
of a spike from cell $a$ at time $t_1^a$ and a spike from cell $b$ at
time $t_2^b$ as $t_1^at_2^b$, the conditional response probabilities
are, to second order:
\eqa
P({\bf 0}|s) & = & 1 - \; \sum_{a=1}^C \sum_{t_1^a}
\overline{r}_{a}(t_1^a ; s)\Delta t + {1 \over 2}
\sum_{ab} \sum_{t_1^a} \sum_{t_2^b}
\overline{r}_{a}(t_1^a ; s) \overline{r}_{b}(t_2^b ; s) \left[1 +
\gamma_{ab}(t_1^a , t_2^b ; s)\right] \Delta t^2
\label{prob_0} \nonumber\\
P(t^a_1|s) &= &  \overline{r}_{a}(t_1^a ; s)\Delta t
  - \overline{r}_{a}(t_1^a ; s) \sum_{b=1}^C \sum_{t_2^b}
\overline{r}_{b}(t_2^b ; s)
\left[1 + \gamma_{ab}(t_1^a , t_2^b ; s)\right] \Delta t^2
  \phantom{ppp} a = 1,\cdots,C \label{prob_1} \nonumber\\
P(t_1^a t_2^b|s) &= & \overline{r}_{a}(t_1^a ; s)
  \overline{r}_{b}(t_2^b ; s) \left[1 +
\gamma_{ab}(t_1^a , t_2^b ; s)\right] \Delta t^2 \phantom{ppp} a=
1,\cdots,C, \qquad b = 1,\cdots,C . \label{eq:resp_prob}
\end{eqnarray}
 \noindent  The unconditional response
probabilities are simply $p(\{t_i^a\}) =
\left<p(\{t_i^a\}|s)\right>_s$. Inserting $p(\{t_i^a\})$ into
Eq.~\ref{eq:entropy} and keeping only leading order terms yields
for the first order total entropy
\begin{equation}
T H_t = \frac{1}{\ln 2} \sum_a \sum_{t_1^a} \left<
\overline{r}_{a}(t_1^a ; s) \Delta t \right>_s - \sum_a
\sum_{t_1^a} \left< \overline{r}_{a}(t_1^a ; s) \Delta t \right>_s
\log_2 \left< \overline{r}_{a}(t_1^a ; s) \Delta t \right>_s .
\label{eq:Ht}
\end{equation}
Inserting $p(\{t_i^a\}|s)$ instead yields a similar expression for the
first order noise entropy $T H^{\rm noise}_t$, except with a {\em
single} stimulus average $\left<\cdot\right>_s$ around the entire
second term.
Continuing the expansion, and noting that a factor of 1/2 is
introduced to prevent overcounting of equivalent permutations, the
additional terms up to second order are:
\eqa
\frac{T^2}{2} H_{tt} &=& \frac{1}{2\ln 2} \sum_{ab} \sum_{t_1^a}
\sum_{t_2^b}
\left\{ \left< \overline{r}_{a}(t_1^a ; s)
\overline{r}_{b}(t_2^b ; s) \left[1 +
\gamma_{ab}(t_1^a , t_2^b ; s)\right]\right>_s
 - \left< \overline{r}_{a}(t_1^a ; s) \right>_s \left<
\overline{r}_{b}(t_2^b ; s) \right>_s\right\} \Delta t^2  \nonumber\\
 && + \sum_{ab} \sum_{t_t^a} \sum_{t_2^b} \left<
\overline{r}_{a}(t_1^a ; s)
\overline{r}_{b}(t_2^b ; s) \left[1 +
\gamma_{ab}(t_1^a , t_2^b ; s)\right]\Delta t^2 \right>_s \log_2
\frac{ \left< \overline{r}_{a}(t_1^a ; s) \right>_s }{
\sqrt{\left<\overline{r}_{a}(t_1^a ; s)
\overline{r}_{b}(t_2^b ; s) \left[1 +
\gamma_{ab}(t_1^a , t_2^b ; s)\right] \right>_s} } \\
\frac{T^2}{2} H_{tt}^{\rm noise} &=& \frac{1}{2\ln 2} \sum_{ab} \sum_{t_1^a}
\sum_{t_2^b}
\left< \overline{r}_{a}(t_1^a ; s)
\overline{r}_{b}(t_2^b ; s)
\gamma_{ab}(t_1^a , t_2^b ; s)\right>_s  \Delta t^2  \nonumber\\
 && + \sum_{ab} \sum_{t_t^a} \sum_{t_2^b} \left<
\overline{r}_{a}(t_1^a ; s)
\overline{r}_{b}(t_2^b ; s) \left[1 +
\gamma_{ab}(t_1^a , t_2^b ; s)\right]\Delta t^2 \log_2 \frac{
\overline{r}_{a}(t_1^a ; s) }{ \sqrt{
\overline{r}_{a}(t_1^a ; s)
\overline{r}_{b}(t_2^b ; s) \left[1 +
\gamma_{ab}(t_1^a , t_2^b ; s)\right]}  }\right>_s .
\label{eq:Htt}
\ena
\vspace{-0.5cm}
\begin{multicols}{2}
\noindent The difference between the total and noise
entropies gives the expression for the mutual information detailed in
\cite{tempinfo}.

It has recently been found that correlations, even if independent
of the stimulus identity, can increase the information present in
a neural population \cite{corrinfo,Abb+99}. This applies both to
cross-correlations between the spike trains from different neurons
and to auto-correlations in the spike train from a single neuron
\cite{tempinfo}. The equations derived above add something to the
explanation of this phenomenon provided in \cite{corrinfo}.
Observe that the second order total entropy can be rewritten in a
form which shows that it depends only upon the grand mean firing
rates across stimuli, and upon the correlation coefficient of the
whole spike train, $\Gamma(t_i^a,t_j^b)$ (defined across {\em
all\/} trials rather than those with a given stimulus as for
$\gamma(t_i^a,t_j^b; s)$). Thus, \eqa \frac{T^2}{2} H_{tt} &=&
\frac{\Delta t^2}{2\ln 2} \sum_{ab} \sum_{t_1^a} \sum_{t_2^b} \left<
\overline{r}_{a}(t_1^a ; s) \right>_s \left<
\overline{r}_{b}(t_2^b ; s) \right>_s \\ \nonumber &\times&
\left\{ \Gamma_{ab}(t^a_i,t^b_j) - [1+\Gamma_{ab}(t^a_i,t^b_j)]
\ln [1+\Gamma_{ab}(t^a_i,t^b_j)] \right\} . \label{2ndorderglobal}
\ena It follows that the second order entropy is maximal when
$\Gamma=0$, and non-zero overall correlations in the spike trains
(indicating statistical dependence) always decrease the total
response entropy. $\gamma(s)$ acts on the noise entropy as
$\Gamma$ does upon the total entropy -- it can only decrease the
conditional entropy. The effect of $\gamma(s)$ on the total
entropy is more complex, depending upon the correlation of the
firing {\em across stimuli}. $\gamma(s)$ can be chosen so as to
increase the total entropy (and thus the information, with the
noise entropy fixed), and this increase will be maximal for the
$\gamma(s)$ which lead exactly to $\Gamma=0$. Neuronal or spike
time interaction may therefore eliminate or reduce the effect of
statistical dependencies introduced by other covariations.

The rate and correlation functions in practice must be estimated from
a limited number of experimental trials, which leads to a bias in each
of the entropy components. This bias was corrected for, as described
in \cite{bias}; however, the sampling advantage that will be described
was observed both with this correction, without bias correction, and
with other bias correction approaches such as that used in
\cite{Strong98}.

\vspace{-0.5cm}
\begin{figure}
\narrowtext
\includegraphics[height=8cm, width=8.6cm]{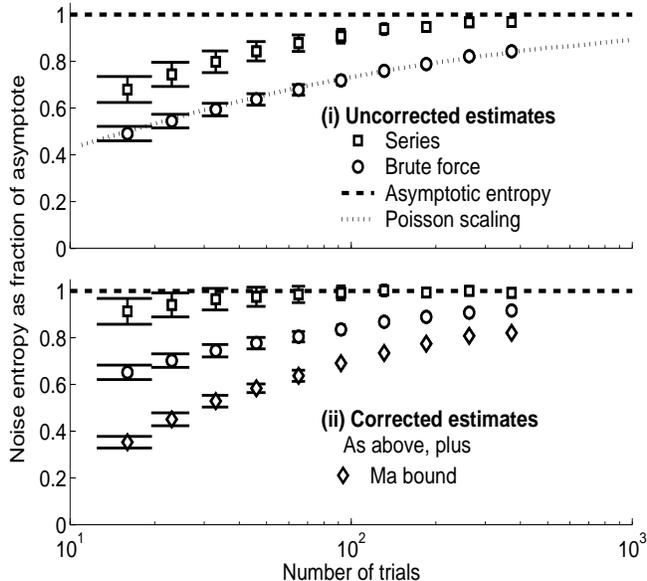}
\caption{Data-size dependence of noise entropy estimates
for a V1 complex cell. Time windows of 40ms (half a stimulus cycle)
were broken into words of length 12 for the analysis. The upper panel,
(i), shows entropy estimates prior to correction for bias,
normalised by the asymptotic (true) entropy. The dotted line indicates
the `brute force' sampling characteristics for a Poisson process with
the same time-dependent firing rate. The lower panel, (ii), shows the
bias-corrected versions of these estimates, and in addition the Ma
lower bound upon the entropy. The asymptotic entropy was obtained by
extrapolating from the curves; the results agree to within
1\%. Error bars were obtained by bootstrap resampling.}
\label{fig:one}
\end{figure}

To demonstrate its applicability, we applied the series entropy
analysis to data recorded from the primary visual cortex (V1) of
anaesthetised macaque monkeys \cite{realdata}. Fig.~\ref{fig:one}
examines, for a typical V1 complex cell, the dependence of the
accuracy of the noise entropy estimate upon the number of experimental
trials utilised. It is the noise entropy which is most affected by
sampling constraints, so we shall concentrate upon this quantity
here. The top panel shows the estimates before application of a bias
removal procedure, using the series (our technique) and `brute force'
(simple application of Eqn.~\ref{eq:entropy}) approaches. The
entropies are expressed as a fraction of the asymptotic entropy
obtained by polynomial extrapolation \cite{Strong98}. 
Reliable extrapolation to the asymptotic entropy was possible because
of the large amount of data that happened to be available for this
cell (which was chosen with that in mind; more usually between 20 and
100 trials were available). This allowed us to compare the performance
of the methods on smaller subsets of the data against a known
reference. The fact that series and brute-force estimators converged
for this cell indicates that higher order correlations amongst spike
times contributed little to the entropy.

The better performance of the series
approach can be understood by considering that (at second order) it
requires sampling from only the first two moments of the probability
distribution, whereas the `brute force' approach depends upon all
moments. Higher moments have to be computed from events with lower and
lower probability, as shown in Eqn.~4; estimation of these lower
probability events is more error-prone, and leads to the larger bias
of the `brute force' approach.

Also shown in Fig.~\ref{fig:one} is the Ma lower bound upon the
entropy \cite{Ma81}, which has been proposed as a useful bound
which is relatively insensitive to sampling problems
\cite{Strong98}. The Ma bound is tight only when the probability
distribution of words at fixed spike count is close to uniform. It
can be seen that for the V1 complex cell data, the Ma bound is not
tight at all. To understand the behaviour of the Ma bound for
short time windows, we calculated series terms. The Ma entropy
already differs from the true entropy at first order: \eqa T
H^{\rm Ma}_t & = & \frac{1}{\ln 2} \sum_a \sum_{t_1^a} \left<
\overline{r}_{a}(t_1^a ; s) \Delta t \right>_s \nonumber\\
&-& \sum_{a;t_1^a}
\left< \overline{r}_{a}(t_1^a ; s) \Delta t \right>_s
\log_2{ \sum_{a ; t_1^a}  \left<
\overline{r}_{a}(t_1^a ; s)\right>_s^2 \Delta t
\over
\sum_{a; t_1^a}  \left<
\overline{r}_{a}(t_1^a ; s)\right>_s
 }
\ena This coincides with Eqn.~5 only if there are no variations of
rate across time and cells. If there were higher frequency rate
variations, or more cells with different response profiles, the Ma
bound would be still less useful.

Estimation quality depends upon not just sampling bias, but also
variance; these can be summarised by the RMS error of the entropy
estimate. We investigated the behaviour of the RMS error by fitting a
Poisson model with matched time-dependent firing rate to the
experimental data of Fig.~1. This model, although yielding a 5\% lower
noise entropy (because of correlations in the real data), predicted
the `brute force' sampling characteristics of Fig.~\ref{fig:one}
almost exactly. The model was used to generate a larger set of data
(10,000 trials, or 160,000 stimulus presentations in total).  This
model yields worst-case sampling for the `brute force' estimator;
worst-case sampling for the series estimator would be achieved by even
spread of probability throughout only the second order response
space. The simulation serves to compare the estimators in a
statistical regime similar to that of the typical cell of
Fig.~\ref{fig:one}.

\vspace{-0.3cm}
\begin{figure}
\narrowtext
\includegraphics[height=9.4cm, width=8cm]{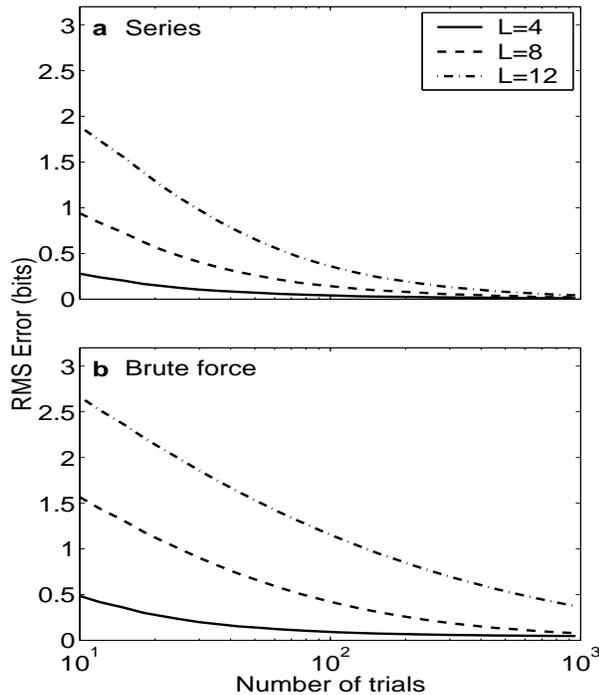}
\caption{RMS error scaling characteristics for wordlengths from 4 to 12 in the simulation. The true noise entropies were 2.0, 3.5 and 4.7 bits respectively.}
\label{fig:two}
\end{figure}

Fig.~\ref{fig:two} shows the scaling of the RMS error before bias
correction with data-size in this simulation. Scaling is
qualitatively similar (but with a sharper decrease) after
correction. The scaling behaviour resulting from the simulation
predicts that with a `brute force' approach, a RMS error of 2\% of
the entropy at a wordlength of 12 would require around 1400 trials
with, and greater than 5000 trials without, application of the
finite sampling correction. The series estimator reduces these
requirements to approximately 50 and 400 trials respectively.
These figures are dependent upon data statistics, and should be
checked on a case by case basis; however, the dimensionality
reduction with the series expansion provides a general improvement
in the quality of entropy estimates for short time windows.

Some readers may wonder whether this new method amounts to computing
the entropy with words with greater than 2 spikes thrown out. This is
not the case: the proposed method considers pairwise interactions
amongst all spikes in the word, no matter how many there are. It thus
(unlike a truncated brute force approach) obtains the ability to take
into account almost all of the entropy of longer words, while
retaining the sampling benefits of being a second order method.

As neuroscience enters a quantitative phase, information theoretic
techniques are being found useful for the analysis of data from
physiological experiments. The methods developed here may broaden the
scope of the study of neuronal information properties. In particular,
they render feasible the information theoretic analysis of some
recordings from anaesthetised and awake mammalian cerebral cortices.

SRS is supported by the HHMI, and SP by the Wellcome Trust.


\begin{thebibliography}{1}
\bibitem{ref:gamma}
The scaled correlation function is measured
as \cite{corrinfo,tempinfo}:
\eqa
\gamma_{ab}(t^a_i,t^b_j;s) &=& \frac{\overline{r_{a}(t_i^a;s)
r_{b}(t_j^b;s)}}{\overline{r}_{a}(t_i^a;
s)\overline{r}_{b}(t_j^b ; s)} -1 , \;\; a \neq b ~ {\rm
or} ~ t_i^a \neq t_j^b \nonumber \\
 \gamma_{aa}(t^a_i,t^a_i;s) &=& -1 \nonumber.
\ena
\end{thebibliography}

\end{multicols}
\end{document}